\documentclass{svjour3}                     
\smartqed  
\usepackage{graphicx}
\newcommand{\beq}{\begin{equation}}
\newcommand{\eeq}{\end{equation}} 
\newcommand{\bea}{\begin{eqnarray}}
\newcommand{\eea}{\end{eqnarray}}

\journalname{Few-Body Systems (FB20)}
\begin{document}

\title{
On the origin and movement of the poles in the coupled channels model for $\bar{K}N$ interactions
\thanks{The work of A.~Ciepl\'{y} was supported by the Grant Agency of Czech Republic, Grant No.~P203/12/2126.}
\thanks{Presented at the 20th International IUPAP Conference on Few-Body Problems in Physics, 20 - 25 August, 2012, Fukuoka, Japan}
}


\author{A.~Ciepl\'{y}         \and
        J.~Smejkal 
}


\institute{A.~Ciepl\'{y} \at
              Nuclear Physics Institute, 250 68 \v{R}e\v{z}, Czech Republic \\
              Tel.: +420-266173284\\
              Fax: +420-220940165\\
              \email{cieply@ujf.cas.cz}           
           \and
           J.~Smejkal \at
              Institute of Experimental and Applied Physics, Czech Technical University in Prague,\\
              Horsk\'{a} 3a/22, 128~00~Praha~2, Czech Republic
}

\date{Received: date / Accepted: date}

\maketitle

\begin{abstract}
The origin of the poles generated by chiral coupled-channel dynamics and applied to $\bar{K}N$ 
interactions is related to the nonzero diagonal couplings in the $\pi\Sigma$, $\bar{K}N$ and $K\Xi$ 
channels. An evolution of the poles from the zero coupling limit and from a limit of restored 
SU(3) symmetry is discussed.

\end{abstract}

The $\bar{K}N$ interaction represents a strongly interacting multichannel system with 
an isoscalar s-wave resonance, the $\Lambda(1405)$, observed in the $\pi\Sigma$ mass 
spectra just below the $\bar{K}N$ threshold. A modern theoretical treatment 
of low energy $\bar{K}N$ interaction is based on chiral perturbation theory combined 
with multichannel techniques that allow to sum properly a major part of the perturbation 
series. The model leads to two dynamically generated resonances 
assigned to the $\Lambda(1405)$, each of them coupling individualy to the $\bar{K}N$ 
and $\pi\Sigma$ states \cite{2003JOORM}, \cite{2008HW}.

It is well known that resonances observed in reaction cross sections and in the transition
amplitudes can be related to the poles of the scattering S-matrix on unphysical Riemann sheets (RS). 
In this short report we concentrate on the pole content of the chirally motivated models 
used for the $\bar{K}N$ interactions and show that only a limited number of poles can 
be generated dynamically within the model framework. Our task is achieved by following 
movements of the poles into (and from) the zero coupling limit (ZCL) in which the inter-channel couplings 
are switched off. We also discuss a limit of a restored SU(3) symmetry (LRSU3) and movements 
of the poles from their SU(3) symmetric states to the physical positions. 

In our analysis we employ the most simple chirally motivated separable model of meson-baryon 
interactions that restricts the inter-channel couplings to the dominant Tomozawa-Weinberg term. 
The model (denoted as TW1) was fitted in Ref.~\cite{2012CS} to the $K^{-}p$ reaction 
and scattering data including the recent SIDDHARTA measurement of the kaonic hydrogen 
characteristics \cite{2011SIDD}. The coupled channels space consists of the $\pi\Lambda$, $\pi\Sigma$, 
$\bar{K}N$, $\eta\Lambda$, $\eta\Sigma$ and $K\Xi$ meson-baryon states that are considered 
(depending on what is calculated) either with proper particle charges or projected to states 
of appropriate isospin. The transition amplitudes matrix $F_{ij}$ (the indexes run over the channel 
space) has poles for complex energies $z$ (equal to the meson-baryon CMS energy $\sqrt{s}$ 
on the real axis) if a determinant of the inverse matrix is equal to zero, 
\beq
{\rm det}|F^{-1}(z)| = {\rm det}|V^{-1}(z) - G(z)| = 0  \;\;\; ,
\label{eq:det}
\eeq
where $V$ stands for the potential matrix, that is proportional to a coupling matrix $C_{ij}$ 
defined by the underlying chiral symmetry, and the intermediate state Green functions 
are represented by the diagonal $G$ matrix. In the hypothetical ZCL, 
in which the non-diagonal inter-channel couplings are switched off ($V_{i,j} = 0$ for $i \neq j$), 
the condition for a pole of the amplitude becomes
\beq
\prod_n [1/V_{nn}(z) - G_{n}(z)] = 0  \;\;\; .
\label{eq:ZCL}
\eeq
There will be a pole in channel $n$ at a RS [$+/-$] (physical/unphysical) 
if the pertinent $n$-th factor of the product on the r.h.s.~of Eq.~(\ref{eq:ZCL}) 
equals zero. Thus, only states with nonzero diagonal couplings $C_{i,j=i} \neq 0$ can generate 
poles in the zero coupling limit. The structure of the SU(3) coefficients that define 
the Tomozawa-Weinberg couplings implies nonzero matrix elements $V_{i,j=i}$ for the $\pi\Sigma$, 
$\bar{K}N$ and $K\Xi$ channels in both isospin sectors, $I=0$ and $I=1$. For each of these channels 
the Eq.~(\ref{eq:ZCL}) has more than one solution, though some solutions are not relevant for physics 
and appear to have only mathematical meaning. In the Table \ref{tab:ZCL} we list the solutions 
that are most relevant for physics for each of the three channels with nonzero diagonal coupling.

\begin{table}
\caption{The pole positions in the zero coupling limit are presented in a form that shows 
their complex energy $z$ in round brackets and the Riemann sheet the pole is found on 
in the square brackets. The characters of the states corresponding to the poles are specified as well.}
\begin{center}
\begin{tabular}{c|lc|lc}
\hline\noalign{\smallskip}
 & \multicolumn{2}{c|}{$I=0$} & \multicolumn{2}{c}{$I=1$}                \\
  channel   & $z$(MeV)[$+/-$] &  status    & $z$(MeV)[$+/-$]  & status   \\ 
\noalign{\smallskip}\hline\noalign{\smallskip}
$\pi\Sigma$ &  (1366,  -91)[-] & resonance &  (1387, -220)[-] & resonance \\ 
$\bar{K}N $ &  (1434,    0)[+] & bound     &  (1158,  -18)[-] & resonance \\
$K\Xi$      &  (1809,    0)[+] & bound     &  (1658,    0)[-] & virtual   \\ 
\noalign{\smallskip}\hline
\end{tabular}
\end{center}
\label{tab:ZCL}
\end{table}

\begin{figure}
\includegraphics[width=0.48\textwidth]{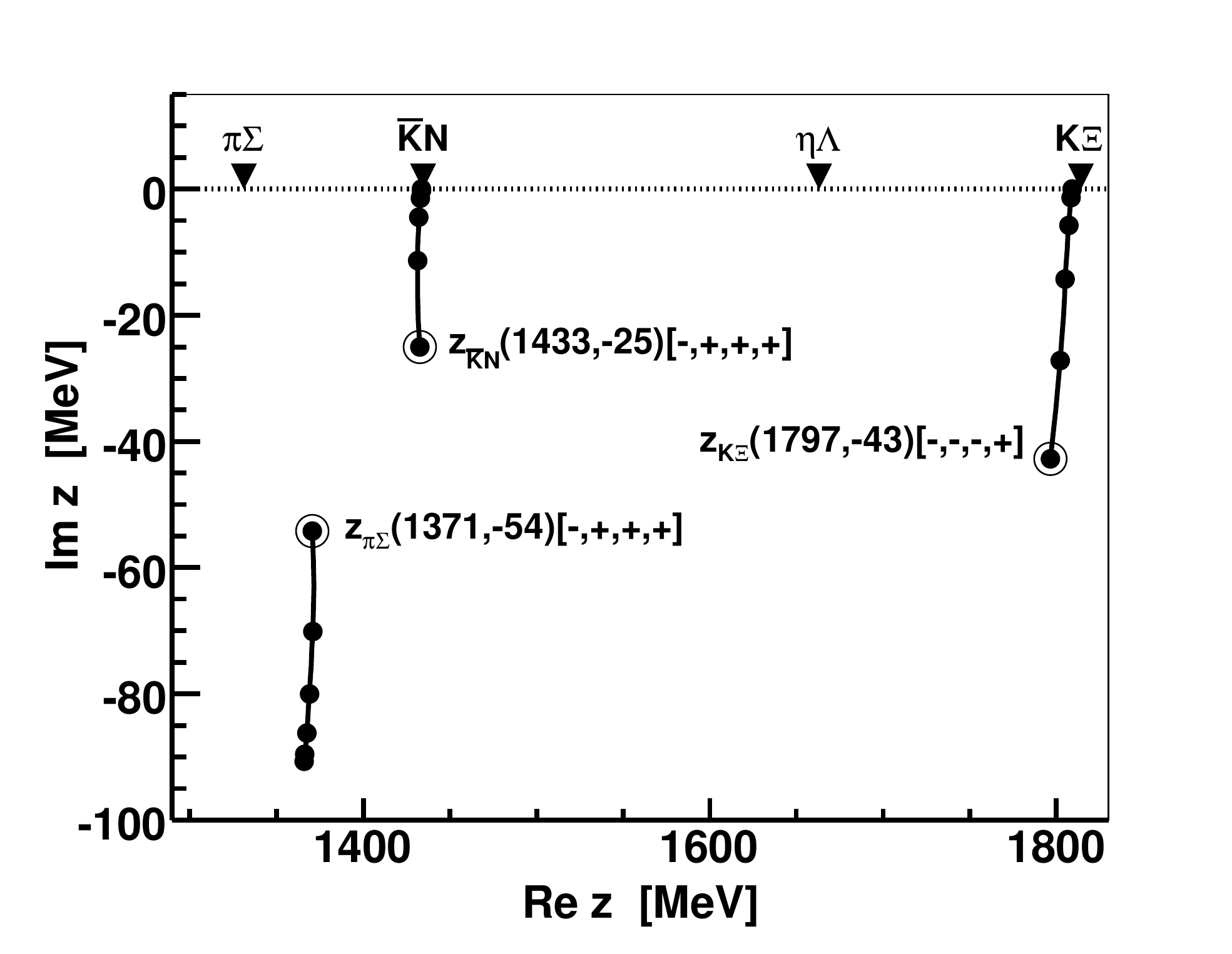} 
\includegraphics[width=0.48\textwidth]{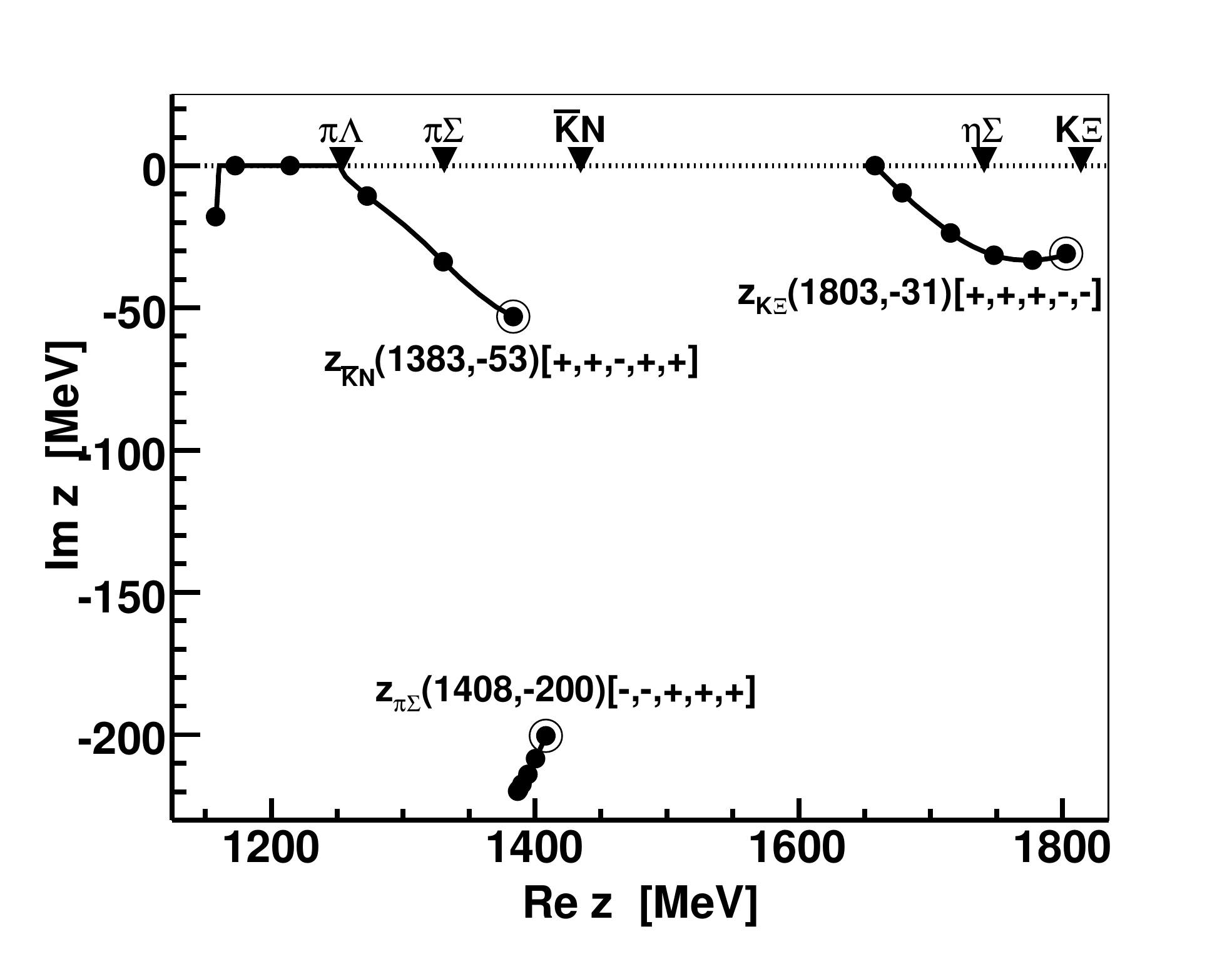} 
\caption{The trajectories of the poles related to the $\pi\Sigma$, $\bar{K}N$ and $K\Xi$ channels 
and obtained by scaling the non-diagonal inter-channel couplings. The left (right) panel relates 
to the isoscalar (isovector) channels. The pole positions in the physical limit are emphasized 
with large empty circles. The triangles at the top of the real axis indicate the channel thresholds.}
\label{fig:ZCL}       
\end{figure}

Once we find the pole positions in the ZCL we can follow their movements on the complex energy 
manifold by gradually turning on the inter-channel couplings. We have done it by scaling 
the non-diagonal couplings by a factor $x$ that ranges from $0$ in the ZCL to $1$ 
in the physical limit. Since the scattering matrix is analytical in $x$ no pole may disappear 
when evolving from its position found for $x=0$. In the multiple channel setup each pole found in the ZCL 
generates a multiplet of shadow poles evolving from the same original position at various RS. 
The pole that is nearest to the physical region plays a dominant role in terms of having an impact 
on physical observables though it may happen that two (or even more) shadow poles are about equally 
away from the physical RS and there is no way to say which of them is dominant. The trajectories 
of the poles that are dominant in the physical limit are visualized in Figure~\ref{fig:ZCL}. 
We have assigned the observed poles to the channels in which they persist in the ZCL and also 
specify the RS 
\footnote{We adopt a notation of the Riemann sheets as strings of $+/-$ signs that match those 
of the imaginary parts of the CMS momenta in all involved channels that are ordered according 
their threshold energies. There are four channels for $I=0$ and five channels for $I=1$ with 
the physical RS denoted as $[+,+,+,+]$ and $[+,+,+,+,+]$, respectively.} 
on which the pole evolves and its final complex energy position in the physical limit. In the ZCL 
the pole positions are those as given in Table~\ref{tab:ZCL}. Of course, the exact positions 
of the poles for $x=0$ and $x=1$ vary with the specific model but the picture is not altered 
qualitatively \cite{2012H}. Specifically, we mention that the ZCL position of the $I=1$ $\bar{K}N$ 
related pole can be brought to the real axis making it a virtual state, e.g. for the next-to-leading 
order NLO30 model specified in Ref.~\cite{2012CS}.  

The isoscalar poles that evolve from the $\pi\Sigma$ resonance and from the $\bar{K}N$ bound state for $x=0$ 
are the two poles that are standardly related to the $\Lambda(1405)$ resonance \cite{2008HW}. In the physical limit 
(for $x=1$) both poles appear on the RS that can be reached from the physical region 
by crossing the real axis in-between the $\pi\Sigma$ and $\bar{K}N$ thresholds. The third isoscalar 
pole $z_{K\Xi}$ can be related to the $\Lambda(1670)$ resonance, though its position is quite off 
the one observed in experimental spectra. We should stress that our TW1 model is not expected to work 
so well at energies hundreds MeV away from the $\bar{K}N$ threshold, so the discrepancy is of little 
concern here. The poles found in the isovector sector are not so well known, only the $\bar{K}N$ one 
was observed earlier too \cite{2003JOORM}. It is understood that it relates to the cusp structure 
in the energy dependence of the elastic $K^{-}n$ amplitude \cite{2010CS}. The $\pi\Sigma$ isovector 
pole is too far from the real axis, so it cannot affect physical observables. The $K\Xi$ 
isovector pole may be related to $\Sigma(1750)$, a three star resonance.

A similar methodology can be applied to study the pole movements from the LRSU3. 
Here we draw our inspiration from Ref.~\cite{2003JOORM} in which the authors discussed a movement 
of the poles into the SU(3) restoration limit upon scaling the meson and baryon masses. 
In this approach the mesons and baryons are assumed to attain channel independent masses $m_{0}=370$~MeV 
and $M_{0}=1150$~MeV, respectively, when the SU(3) symmetry is restored.
The SU(3) decomposition of the meson and baryon octets combination provides us with a SU(3) singlet 
state, symmetric octet, antisymmetric octet, decuplet, antidecuplet and 27-plet states. In the SU(3) 
eigenstates basis the coupling matrix is diagonal, $\tilde{C}_{ij} = {\rm diag}(6,3,3,0,0,-2)$, 
where the eigenvalues are in the pertinent order. Thus, the condition for pole existence in the factorized 
form of Eq.~(\ref{eq:ZCL}) applies to the LRSU3 with the index $n$ running 
over the listed SU(3) eigenstates and with the effective potentials $V_{ij}$ replaced by a diagonal 
potential matrix proportional to the $\tilde{C}$ couplings. The solutions of Eq.~(\ref{eq:ZCL}) 
applied to the LRSU3 are given in Table~\ref{tab:SU3}. Two solutions are 
found for each SU(3) eigenstate, all of them on the real axis. Since the $\tilde{C}$ 
eigenvalues of the symmetric and antisymmetric octets are degenerate only one set of solutions is shown 
for a generic octet SU(3) eigenstate. We also note that some solutions are of purely mathematical 
character and do not have a good physical meaning. This applies specifically to the 27-plet solutions 
as the negative value of the pertinent coupling indicates a repulsive character of the 27-plet force.

\begin{table}
\caption{The pole positions in the SU(3) restoration limit. Only real parts of the complex energies 
$z$ are shown (with Im $z\;=\;0$ in all cases) followed by a specification of the Riemann sheet
the pole is found on.}
\begin{center}
\begin{tabular}{c|cc|cc|cc|}
\hline\noalign{\smallskip}
SU(3) state   & \multicolumn{2}{c|}{singlet} & \multicolumn{2}{c|}{octet}  & \multicolumn{2}{c|}{27-plet} \\ 
eigenvalue    & \multicolumn{2}{c|}{$6$}     & \multicolumn{2}{c|}{$3$}    & \multicolumn{2}{c|}{$-2$}    \\
\noalign{\smallskip}\hline\noalign{\smallskip}
Re $z$(MeV)[$+/-$] & 1456[+] &  1188[-]      &    1518[-]     &  1235[-]   &   1056[-]     & 784[+]       \\ 
status        &    bound   &   virtual       &    virtual     &  virtual   &   virtual     & bound        \\ 
\noalign{\smallskip}\hline
\end{tabular}
\end{center}
\label{tab:SU3}
\end{table}

The movement of the poles (upon scaling the meson and baryon masses) from their positions in the physical 
limit to the SU(3) singlet and octet states was discussed in Refs.~\cite{2003JOORM} and \cite{2012H}. 
In those works only the pole positions at higher energies were considered. One of the isovector poles 
shown in Figure~\ref{fig:ZCL} was also missing in \cite{2003JOORM} and another one disappeared 
on its way from the SU(3) octet state to the physical limit. With our methodology we are able to follow 
the whole trajectories of all poles including those that go to (or start from) the SU(3) eigenstates 
realized as poles at lower energies. We were able to relate the differences between our results and those 
of Ref.~\cite{2003JOORM} to a different treatment of RS. A detailed account on the pole 
movements to the LRSU3 and related issues is being prepared for publication. Here we just 
summarize that the LRSU3 provides a richer pole content than anticipated earlier. 
The main merit of the present work lies in tracking the origin of the poles generated dynamically 
within the coupled channel chiral models of $\bar{K}N$ interactions to those realized in the ZCL. We also 
maintain that there are at least three isoscalar poles and three isovector poles arising from 
the $\pi\Sigma$, $\bar{K}N$ and $K\Xi$ diagonal interactions.


\vspace*{-2mm}
\begin{thebibliography}{3}
%
\vspace*{-3mm}
\bibitem{2003JOORM}
  Jido, D., Oller, J.\,A., Oset, E., Ramos, A., Mei{\ss}ner, U.-G.,
  Nucl.~Phys.~A~725, 181 (2003)
\bibitem{2008HW}
  T.~Hyodo, W.~Weise,
  Phys.~Rev.~C 77, 035204 (2008)
\bibitem{2012CS} 
  Ciepl\'{y}, A., Smejkal, J., 
  Nucl.~Phys.~A 881, 115 (2012)
\bibitem{2011SIDD}
  Bazzi, M. {\it et al.} [SIDDHARTA Collaboration], 
  Phys.~Lett.~B 704, 113 (2011)
\bibitem{2012H}
  Hrazdilov\'{a}, L., 
  graduate thesis, Czech Technical University (2012)
\bibitem{2010CS}
  Ciepl\'{y}, A., Smejkal, J.,
  Eur.~Phys.~J.~A 43, 191 (2010)

%
\end{thebibliography}
\end{document}